\documentclass[oldversion, referee]{aa}
\usepackage{times,graphicx,amssymb,lscape}
\usepackage{rotating}
\usepackage{natbib}
\bibpunct{(}{)}{;}{a}{}{,}  % to follow AA style

\begin{document}

\title{The periodic variations of a white--light flare observed with ULTRACAM}

\author{M. Mathioudakis\inst{1}
        \and
        D.S. Bloomfield\inst{1}
	\and
        D.B. Jess\inst{1} 
        \and
        V.S. Dhillon\inst{2}
        \and
        T.R. Marsh\inst{3}
        }

\institute{Department of Physics and Astronomy, Queen's University Belfast,
	Belfast, BT7~1NN, Northern Ireland, UK 
\and
Department of Physics and Astronomy, University of 
Sheffield, Sheffield, S3 7RH, UK
\and
Department of Physics, University of Warwick, Coventry, 
CV4 7AL, UK}

\offprints{M. Mathioudakis, \email{M.Mathioudakis@qub.ac.uk}}

\date{Received date / Accepted date}

\abstract{
 High time resolution observations of a white--light flare on the active 
star EQ PegB show evidence of intensity variations with a period of 
$\approx$10~s. The period drifts to longer values during the decay phase of 
the flare. If the oscillation is interpreted as an impulsively--excited, 
standing--acoustic wave in a flare loop, the period implies a loop length of 
$\approx$1.7~Mm and $\approx$3.4~Mm for the case of the fundamental 
mode and the second harmonic, respectively. However, the small loop 
lengths imply a very high modulation depth making the acoustic interpretation 
unlikely.  A more realistic interpretation may be that of a fast--MHD wave, 
with the modulation of the emission being due to the magnetic field. 
Alternatively, the variations 
could be due to a series of reconnection events. The periodic signature may 
then arise as a result of the lateral separation of individual flare loops 
or current sheets with oscillatory dynamics ({\emph{i.e.}}, periodic 
reconnection).

%%\end{abstract}
}

\authorrunning{M. Mathioudakis et~al.}
\titlerunning{Flare oscillations}

\maketitle 
 
\section{Introduction}
%\label{intro}
The detection of oscillations in coronal loops has provided clear evidence 
for waves in the upper solar atmosphere (Nakariakov \& Verwichte 2005). These 
observations provide new insights into the processes of atmospheric 
heating. They have also raised the prospect of using oscillations as a 
diagnostic to infer the properties of the upper solar atmosphere and 
allowed the development of coronal seismology (Roberts 2000). For example, 
Nakariakov \& Ofman (2001) developed a method for determining an absolute 
value for the magnetic field strength of coronal loops that oscillate as a 
result of standing kink--mode waves. The crucial 
theory of MHD oscillations in solar atmospheric structures has been 
presented by Roberts, Edwin, \& Benz (1984), where both standing and 
propagating wave modes are considered. Coronal structures, such as the 
loops within an active region, can act as wave guides supporting 
quasi--periodic variations (Roberts, Edwin, \& Benz 1983).

There is increasing evidence to suggest that oscillations in the solar 
atmosphere can be triggered by flares and other nearby impulsive events. 
Kane et al. (1983) observed 8~s pulsations of large amplitude during the 
hard X-ray (HXR) and microwave bursts of a solar flare, while McKenzie \& 
Mullan (1997) have reported periods of $10 - 60$~s in the active, 
non--flaring solar corona. Schrijver, Aschwanden, \& Title (2002) have 
shown that oscillation events observed by the {\emph{Transition Region and 
Coronal Explorer}} ({\sl{TRACE}}) are often triggered by flares and 
filament eruptions, occurring in closed coronal loops with no clear 
dependence of the oscillation amplitude on the magnitude of the flare. 
Lower in the atmosphere, McAteer et al. (2005) have shown that a flare 
induced an oscillatory signal along an H$\alpha$ ribbon. 

The solar analogy is often used to interpret the wide range of phenomena 
observed in active stars. Quasi--periodic fluctuations ($\approx$30~s) of 
varying amplitude were reported in the quiescent state of active flare 
stars and were considered to be a signature of microvariability (Andrews 
1989). A similar study by Mullan, Herr, \& Bhattacharyya (1992) reported 
periodicities of a few minutes which were interpreted as transient 
oscillations in coronal loops. There are also some reports of periodic 
intensity variations during stellar flares. Short duration variations 
($10 - 20$~s) were reported after the onset of a strong flare on 
the Hyades star II Tau (Rodon\'{o} 1976), while Mathioudakis et al. (2003) 
reported a large amplitude periodicity (220~s) during the peak of a flare 
on the RS CVn binary II Peg. The first stellar flare oscillation in X--rays 
was recently reported by Mitra--Kraev et al. (2005).

Quasi--periodic oscillations during flares can be generated by both the 
second harmonic of the acoustic mode 
(Nakariakov et al. 2004, Tsiklauri et al. 2004) and the fundamental mode 
(Taroyan et al. 2005). The numerical modeling shows that the loop density 
exhibits perturbations with a maximum near the loop apex, with the 
oscillation period dependent on the ratio of the loop length to the sound 
speed (Roberts, Edwin, \& Benz 1984). This approach is particularly useful 
when applied to stellar flares as it provides a method for determining the 
spatial dimensions of stellar coronal loops.    

EQ Peg (Gl 896AB) is a visual binary system with a separation of 5$''$. Both 
components are M--type flare stars with visual magnitudes of 10.3 and 
12.4, respectively. Flare activity on the system has been observed over a 
large part of the electromagnetic spectrum from X--rays to radio 
wavelengths. In white light, the flare frequency of the binary is 
$\sim$0.8~flares hr$^{-1}$ for flares with energy in excess of 10$^{30}$~ergs 
(Lacy, Moffett, \& Evans 1976).

Small--amplitude stellar variations are often met with some skepticism as 
they may be due to sky fluctuations. To bypass this difficulty  multi--site 
observations may be used (Zhilyaev et al. 2000). Here we 
use high--cadence, multi-wavelength photometry to study periodic intensity 
variations during a flare on EQ PegB. Synchronous observations of a 
comparison star were obtained in order to remove any atmospheric 
variations. Section~2 details the instrumentation and 
observations, with the form of time series analysis presented in Section~3. 
The results are discussed in Section~4, while our conclusions are drawn in Section~5.

\section{Observations and data reduction}
\label{obs_analy}
 The observations presented here were obtained with the triple--beam CCD 
instrument {\sc{ultracam}} (Dhillon \& Marsh 2001). The instrument uses two 
dichroic beam--splitters to separate the light into three wavelengths 
which pass through differing Sloan ({\sc{sdss}}) filters. The detector 
system comprises of three back--illuminated Marconi 1024x1024 
frame--transfer CCDs, allowing the collection of data 
in the imaging area while data in the masked area is read out. Each chip is 
triggered by the same GPS synchronized system clock.

We used {\sc{ultracam}} on November 4$^{\mathrm{th}}$ 2003 with the 4.2~m 
William Herschel Telescope on La Palma, resulting in a spatial sampling of 
0.3$''$~pixel$^{-1}$. The large imaging area of the CCDs allows simultaneous 
observations of the target and a sufficiently bright comparison star.  
Smaller windows of 21$''$x21$''$ were used around these objects to minimize the 
dead time between exposures. The observations were made in the $u'$, $g'$, 
and $r'$ Sloan passbands with an exposure time of $\sim$0.073~s. 
The data were reduced with the {\sc{ultracam}} pipeline reduction 
software. The pipeline uses aperture photometry to produce separate 
lightcurves for each selected object within the field of view 
({\emph{i.e.}}, both the target and the comparison).
%Figure 1

\begin{figure}
\begin{center}
\includegraphics[angle=90,width=8.75cm]{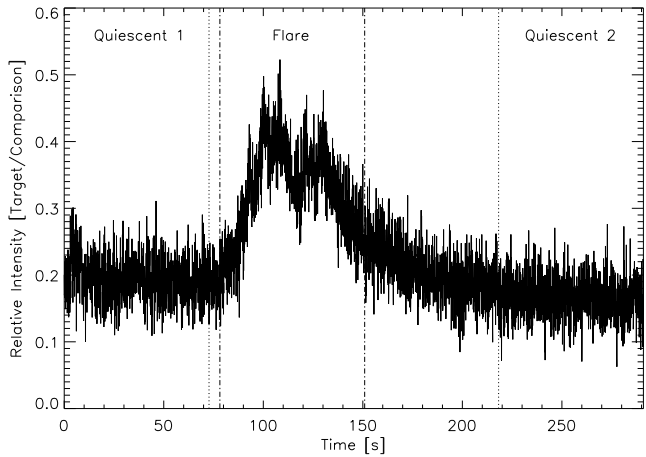}
\caption{The $u'$--band lightcurve of the analyzed flare on EQ PegB. Time 
is given in seconds since 21:32:53~UT on November 4$^{\mathrm{th}}$ 2003.}
\label{fig:time_flare}
\end{center}
\end{figure}

\section{Time series analysis}
 The $u'$--band lightcurve of the flare on EQ PegB is shown in 
Figure~\ref{fig:time_flare}, where the duration of the flare is 
$\approx$160~s. The impulsive phase of the event is followed by intensity 
variations which were analyzed by both a Fast Fourier Transform and a 
wavelet transform. Although wavelet analysis is similar to a windowed 
Fourier analysis, the latter may be considered as an inefficient method for 
the study of quasi--periodic oscillations. This arises not only from 
Fourier analysis comparing different numbers of oscillatory cycles for 
differing analysis frequencies, but the windowing process imposes a shorter 
time interval on the time series within which the analysis is carried out. 
Hence, the frequency resolution of the Fourier analysis is degraded since 
$\Delta \nu = 1/T$, where $T$ is the analysis duration.

With wavelet analysis the search for periodic signatures is carried out by 
a time--localized function which is continuous in both frequency and time 
(Torrence \& Compo 1998), making it well suited for the identification of 
transient oscillations. Wavelets have become the preferred 
analysis technique in recent years because wave phenomena in solar and stellar 
atmospheres have finite durations. An extensive discussion on the effects 
that various wavelet parameters have upon their results is given in De 
Moortel, Munday, \& Hood (2004).

\begin{figure}[]
\begin{center}
\includegraphics[angle=90,width=8.75cm]{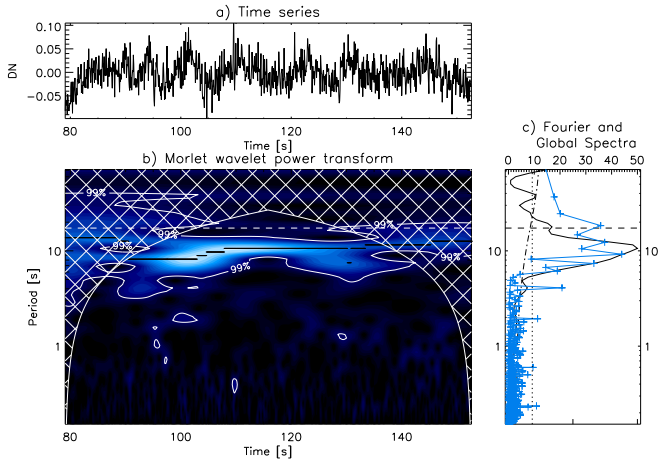}
\caption{{\emph{a}}) The filtered $u'$--band time series. {\emph{b}}) The 
Morlet wavelet power transform of the time series in {\emph{a}}) as a 
function of both time (abscissa) and oscillation period (ordinate). Contours 
define the power above which detections are 99\% confident using a two 
degree of freedom $\chi^2$ distribution. {\emph{c}}) The global 
({\emph{i.e.}}, averaged) wavelet power (full line) and Fast Fourier power 
(crosses) as a function of period. The 99\% confidence levels are 
over--plotted for the global wavelet and Fourier curves as dot--dashed and 
dotted lines, respectively.}
\label{fig:wave_flare}
\end{center}
\end{figure}

The wavelet used in this study is the Morlet function, which is defined as 
the product of a complex exponential ({\emph{i.e.}}, sine) wave with a 
Gaussian envelope,
\begin{equation}
\psi_t \left( s \right) = \pi^{-1/4} \exp \left( i \omega t \right) \exp \left( \frac{-t^2}{2s^2} \right) \ ,
\end{equation}
where $t$ is the time parameter, $s$ is the wavelet scale (related to the 
Fourier period by $P = 1.03s$ for the Morlet function), $\omega$ is the 
oscillation frequency parameter, and $\pi^{-1/4}$ is a normalization term 
(Torrence \& Compo 1998). By varying the scale of the wavelet function 
({\emph{i.e.}}, both the Gaussian width and the oscillation period) such 
that the sinusoidal portion matches a given oscillation frequency, the 
wavelet is convolved with the time series to determine the contribution of 
that frequency to the time series. At each wavelet scale, the timing 
information is achieved by scanning the wavelet function through the 
time series.

The portions of the $u'$, $g'$, and $r'$ time series occurring between the 
vertical dot--dashed lines in Figure~\ref{fig:time_flare} were filtered 
using wavelet reconstruction to remove the long--period power associated 
with the general flare profile. This technique rebuilds the time series, 
$x_t$, from the wavelet transform, $W_t(s)$, using only information from 
the desired scale values, $s_j$,
\begin{equation}
x_t = \frac{\delta j \delta t^{1/2}}{C_{\delta}\psi_0(0)}\sum_{j=0}^{J}\frac{\Re\lbrace W_t(s_j) \rbrace}{s_j^{1/2}} \ ,
\end{equation}
where $\delta j$ defines the degree of frequency sampling (taken as $1/32$ 
here), $\delta t$ is the cadence ($0.073$~s), and both $C_{\delta}$ and 
$\psi_0(0)$ are constants ($0.776$ and ${\pi}^{-1/4}$ for the Morlet 
wavelet, respectively). The time series presented here were reconstructed 
over all scale values corresponding to periods less than 17~s, a point 
discussed in more detail later. The wavelet power transform of our filtered 
$u'$--band time series is shown in Figure~\ref{fig:wave_flare} where 
regions of lighter shading indicate greater oscillatory power. The 
cross--hatched area is the cone of influence (COI) and defines the region 
of the power spectrum where edge effects may become important due to the 
finite duration of the time series --- the extent of the COI at each period 
is the $e$--folding or decorrelation time of the wavelet function.

\begin{figure}[]
\begin{center}
\includegraphics[angle=0,width=8.75cm]{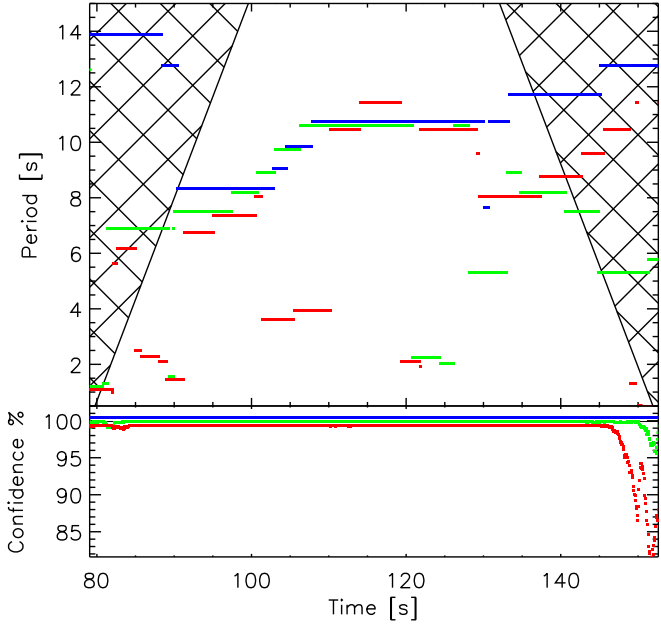}
\caption{{\emph{Upper}}: Oscillation period exhibiting the peak wavelet 
power at each point in time. {\emph{Lower}}: Confidence levels achieved by 
peak wavelet power through randomization testing. In both panels the 
results from the $u'$ (blue), $g'$ (green), and $r'$ (red) bands are 
plotted separately.}
\label{fig:wave_peaks}
\end{center}
\end{figure}

The reliability of any oscillatory power seen in our data is tested against 
a number of criteria:
\begin{enumerate}
\item Power is tested against spurious detections which could be due to 
Poisson noise. The contours in Figure~\ref{fig:wave_flare}{\emph{b}} 
outline the power at which detections are 99\% confident using a two degree 
of freedom $\chi^2$ distribution.
\item Time series are compared to a large number of randomized series 
(1500) with identical count statistics. If there are no periodic signals in 
the data the measured peak power values should not depend on their 
observation times. A random--detection probability, $p$, is calculated for 
the peak wavelet power at each point in time by comparing the number of 
times that the random series produce equal or greater power than the actual 
data --- high values indicating no periodic signals in the data while low 
values suggest the detected peak power periods are real. Confidence levels 
are calculated from $(1 - p) \times 100$ and are shown for each of our time 
series in the lower panel of Figure~\ref{fig:wave_peaks}.
\item As mentioned previously, the extent of the COI is the decorrelation 
time of the wavelet function. For the Morlet wavelet this is $\sqrt{2}P$, 
where $P$ is the oscillatory period. The choice of cutoff period for the 
filtering (17~s) was based on this value, as longer periods ({\emph{i.e.}}, 
those above the dashed line in Figure~\ref{fig:wave_flare}{\emph{b}}) can 
not exist for a decorrelation duration outside the COI. To minimize the 
chance of detecting noise spikes, we require real detections to have power 
above the 99\% confidence level for at least three oscillatory cycles 
({\emph{i.e.}}, two decorrelation durations).
\end{enumerate}

\section{Results and discussion}
Our analysis shows a reliable oscillation with a period of $\approx$10~s 
during the flare on EQ PegB. The oscillation starts just before flare 
maximum ($t \approx 100$~s) and continues into the decay phase. We conclude 
that it is due to white--light continuum variations rather than the Balmer and 
Ca {\sc{ii}} H\&K lines which can provide a significant contribution to the 
$u'$--band flux during a flare, because it is seen in all three bands ($u'$, 
$g'$, and $r'$). The following interpretations may be considered.

\subsection{An impulsively--generated acoustic wave}
The generation of standing, slow--mode waves in active, coronal loops which 
are anchored in the photosphere was first proposed by Roberts, Edwin, \& 
Benz (1984). Assuming both a strong longitudinal component (in order to 
yield density and hence intensity variations) and that the magnetic field 
dominates sufficiently to approximate the longitudinal tube speed with the 
acoustic speed, the period of a standing, slow--mode oscillation is given by,
\begin{equation}
\label{eqn:standing}
P \left( s \right) \approx \frac{ 2 L \left( Mm \right) }{ 7.6 \times 10^{-2} N \sqrt{ T \left( MK \right) } } \ ,
\end{equation}
where $T$ is the average temperature along the loop, $L$ is the length of 
the loop, and $N$ is the mode of oscillation ({\emph{i.e.}}, 1 yields the 
fundamental mode, 2 the second harmonic). In Figure~\ref{fig:wave_peaks} we 
show the oscillatory periods exhibiting the maximum power at each point in 
time. In all three {\sc{ultracam}} bands, the period shifts to longer values 
during the decay phase of the flare. This behaviour can be explained by the 
standing, acoustic--wave model as a result of a temperature decrease and/or 
increase in loop length after the flare peak. Substituting the detected 
oscillation period (10~s) and a temperature of 20~MK into 
Equation~\ref{eqn:standing} we obtain a loop length of $\approx$1.7~Mm 
($0.0075 R_\star$) for the fundamental mode and $\approx$3.4~Mm ($0.015 
R_\star$) for the second harmonic.

In a recent numerical model, Nakariakov et al. (2004) have shown that the 
standing second harmonic of an acoustic wave can be excited in a flaring 
loop. The excitation of these oscillations is almost independent of the 
location of the heat deposition in the loop (Tsiklauri et al. 2004) and appears as a 
natural response of the loop to an impulsive energy deposition. Taroyan et 
al. (2005) have also examined the excitation of acoustic waves as a result 
of an impulsive energy deposition at the chromospheric footpoint of a loop. 
They show that different pulses at the footpoint result in oscillations 
with the fundamental--mode period. Only pulses with a duration 
equal to the fundamental--mode period manage to set up standing waves, the 
rest generate propagating waves. 

In these models, the density fluctuations that are generated at the end of
the coronal loops, which are located at the top of the chromosphere, may account
for the variation of the optical emission (see Fig. 2 in Nakariakov et
al. 2004). Alternatively, the optical emission could arise from the free-free 
emission of coronal loops, as suggested by Mullan et al.(1992). Stepanov et al. 
(2005) have criticized the latter interpretation as it would imply a high 
plasma--$\beta$ which would make the loops unstable. However, since we are 
considering a flare loop, which is intrinsically unstable, the Mullan et al. 
(1992) interpretation can not be excluded.

We would like to re--emphasize that one of the main difficulties with the acoustic 
interpretation is the very small loop length. This implies an extremely large 
modulation depth due to the flare site occupying a small fraction of the stellar 
surface. This difficulty has also been pointed out by Stepanov et al. (2005) for a 
flare on EV Lac.  

\subsection{A fast--magnetoacoustic wave --- the sausage mode}
The global sausage mode (GSM) is a fast--MHD mode which can efficiently modulate 
the plasma density and magnetic field strength. It is also one of the principal 
modes that can be excited in coronal loops. The modulation of the
optical emission from the loop footpoints is determined by the flux of the 
high--energy electrons which penetrate the lower atmosphere (Stepanov et al. 2005).
The sausage mode can be supported if the loops are sufficiently thick and 
dense. In order to assess whether the GSM can be supported in this flare we 
require information on the loop length, temperature, density and magnetic field 
strength. Mullan et al. (2006) have shown that EUV and X-ray observations 
provide a powerful tool for estimating the physical parameters in the 
flaring loops of active stars with EQ Peg being one of the stars in their study. 
For this flare we select a temperature $T = 5 \times 10^{7}$~K, an internal loop 
density of $n_e^f = 4 \times 10^{12}$~cm$^{-3}$, a loop length of $L = 1.8 \times 
10^{9}$~cm (0.14$R_{\star}$), and a field strength of $B = 1100$~G (Mullan et al. 
2006). We assume that the quiescent electron density is $n_e^q = 4 \times 
10^{11}$~cm$^{-3}$. The GSM exists if the dimensionless wave number, $k\alpha$, 
is greater than the cut--off value,
\begin{equation}
k_c\alpha = j_0\sqrt{\frac{(C_{S0}^2 + C_{A0}^2)(C_{Ae}^2 - C_{T0}^2)}
{(C_{Ae}^2 -C_{A0}^2)(C_{Ae}^2 - C_{s0}^2)}}
\end{equation}
where $\alpha$ is the loop radius and $j_0$ = 2.4. $C_{A0}$ and $C_{Ae}$ are the 
internal and external Alfv\'{e}n speeds, while $C_{S0}$ and $C_{T0}$ are the 
internal sound and tube speeds, respectively. We will follow the criteria 
outlined in Nakariakov et al. (2003) to determine if this mode can explain the 
oscillations detected in the EQ PegB flare. Using the loop parameters, we 
calculate $C_{A0} = 1,200$~km~s$^{-1}$, $C_{Ae} = 5,560$~km~s$^{-1}$, $C_{s0} = 
1,175$~km~s$^{-1}$ and $C_{T0} = 840$~km~s$^{-1}$. This leads to $k_c\alpha = 
0.75$ and $\alpha = 0.24 L$. The ratio of the Alfv\'{e}n speeds is consistent 
with these parameters ($L/2\alpha < \frac{\pi C_{Ae}}{2 j_0 C_{A0}}$). The phase 
speed $C_p = \frac{2L}{P} = 3,600$~km~s$^{-1}$, which is less than the cut--off 
value ($C_p(k_c) = C_{Ae}$) of the GSM. Finally, the internal Alfv\'{e}n speed 
and loop radius estimated above would imply that the period for the GSM should be 
$P_{GSM} < \frac{2\pi \alpha}{j_0 C_{A0}} = 9.3$~s. We conclude that a fast--MHD 
wave ({\emph{i.e.}}, the sausage mode) could explain the oscillation detected in 
this flare, because the shortest period detected is 8~s (Figure~\ref{fig:wave_peaks}).

\subsection{Individual flare bursts}
A detailed survey of solar white--light flares observed with both 
{\sl{TRACE}} and the {\emph{Reuven Ramaty High Energy Solar Spectroscopic 
Imager}} has confirmed the strong association of white--light emission with 
HXR emission (Hudson, Wolfson, \& Metcalf 2006). Some solar HXR bursts show 
a high degree of periodicity and have been attributed to the periodic 
injection of electron beams into the chromosphere (Aschwanden et al. 1994). 
They suggest that the repetitive injection of beams is governed by a single 
quasi--periodic accelerator rather than a spatially fragmented system.

The intensity variations observed during the flare on EQ PegB could also be 
attributed to a group of individual bursts occurring at $\tau \approx 10$~s 
intervals. Emslie (1981) used an interacting--loop model to explain the time 
structure and periodicity observed in solar--HXR bursts. In his model, the 
impulsive energy released in one flare loop creates a disturbance in the 
surrounding field lines. The disturbance propagates roughly horizontally 
with a speed approximately equal to the Alfv\'{e}n speed, $V_A = B / 
( 4 \pi n_H m_H )^{1/2}$, until it encounters a neighbouring loop at a spatial 
separation of $D \approx V_A \tau$. The second loop is triggered and 
produces a burst which is spatially and temporally different from the first 
one. Spectroscopic measurements of photospheric magnetic field strengths in 
active M stars have revealed values in the range $2,000 - 4,000$~G (Saar \& 
Linsky 1985, Saar et al. 1986). Adopting a value of 3,000~G for EQ PegB and a 
hydrogen number density from the stellar atmosphere models of Hawley \& 
Fisher (1992), we derive a sequential loop separation of $D \approx  1$~Mm 
for the series of loops. The drift to larger $\tau$ implies an increase of 
spatial separation with time during the flare.

\section{Concluding remarks}
\label{conc}
High time resolution, ground--based observations of a white--light flare on 
an active, fully--convective star reveal intensity variations in continuum 
emission with a period of $\approx$10~s. The variations are observed in the 
Sloan $u'$, $g'$, and $r'$ bands using synchronous observations of a 
comparison star to eliminate the possibility that these perturbations are 
due to changes in the Earth's atmosphere. If the variations are interpreted 
as an impulsively--excited, standing--longitudinal wave, the period of the 
oscillation carries information on the physical properties of the medium in 
which it occurs. However, one of the main drawbacks of the acoustic wave 
interpretation is that it predicts a very small loop length, which may be 
unrealistic given the large amount of energy involved during the event. Using 
typical coronal loop parameters for EQ PegB, we show that a fast--MHD wave, 
the sausage mode, could provide a more viable interpretation for the observed 
oscillation. The observations show the potential of applying solar atmospheric 
seismology techniques to stellar studies. 

A series of individual bursts arising from periodic magnetic reconnection could 
also explain the periodic variations observed. Although reconnection is a dynamic 
phenomenon with a high degree of intermittency it may lead to periodic signatures. 
For example, one of the main characteristics of the Tajima et al. (1987) current 
loop coalescence model is the appearance of quasi--periodic oscillations in the 
electric and magnetic field energies and ion temperature. The electric 
field explosively increases and subsequently oscillates as the magnetic 
flux in the coalesced loops is alternatively compressed and decompressed. 
A promising interpretation, suggested recently by Nakariakov et al. (2006), 
involves the MHD oscillation of a non-flaring loop ({\emph{i.e.}}, the driver) 
interacting with a flaring active region. This interaction can lead to periodic 
variations of the current density with a modulation depth significantly higher 
than the driving oscillation.
 
\acknowledgements
This work has been supported by the UK Particle Physics and Astronomy Research 
Council (PPARC). TRM acknowledges the support of a PPARC Senior Research 
Fellowship. ULTRACAM is supported by PPARC grant PPA/G/S/2002/00092. We would 
like to thank Prof. D.J. Mullan and Dr. V. Nakariakov for useful discussion. We 
would also like to thank the referee, Dr. Stepanov, for his comments and 
suggestions on the paper.

%%%%% the bibliography
\bibliography{aa}
\bibliography{submit}

\bibitem[Andrews 1989]{Andrews1989}
Andrews, A.D., 1989, \aap, 214, 220

\bibitem[Aschwanden, Benz, \& Montello 1994]{Aschwanden1994}
Aschwanden, M.J., Benz, A.O., Montello, M.L., 1994, \apj, 431, 432

\bibitem[De Moortel, Munday, \& Hood 2004]{DeMoortel2004}
De Moortel, I., Munday, S.A., Hood, A.W., 2004, \solphys, 222, 203

\bibitem[Dhillon \& Marsh 2001]{Dhillon2001}
Dhillon, V.S., \& Marsh, T.R., 2001, New Ast.Rev., 45, 91

\bibitem[Emslie 1981]{Emslie1981}
Emslie, A.G., 1981, Astrophysical Letters, 22, 41

\bibitem[Hawley \& Fisher]{Hawley1992}
Hawley, S.L., Fisher, G.H., 1992, \apjs, 78, 565

\bibitem[Hudson, Wolfson, \& Metcalf 2006]{Hudson2006}
Hudson, H.S., Wolfson, C.J., Metcalf, T.R, 2006, \solphys, in press

\bibitem[Kane et al. 1983]{Kane1983}
Kane, S.R., et al., 1983, \apj, 271, 376

\bibitem[Lacy, Moffett, \& Evans 1976]{Lacy1976}
Lacy, C.H., Moffett, T.J., Evans, D.S., 1976, \apjs, 30, 85

\bibitem[Mathioudakis et al. 2003]{Mathioudakis2003}
Mathioudakis, M., et al. 2003, \aap, 403, 1101

\bibitem[McAteer et al. 2005]{McAteer2005}
McAteer, R.T.J., et al. 2005, \apj, 620, 1101

\bibitem[McKenzie and Mullan 1997]{McKenzie1997}
McKenzie, D.E., \& Mullan, D.J. 1997, \solphys, 176, 127

\bibitem[Mitra-Kraev et al. 2005]{Mitra2005}
Mitra-Kraev, U., Harra, L.K., Williams, D.R., Kraev, E. 2005, \aap, 436, 1041

\bibitem[Mullan, Herr, \& Bhattacharyya 1992]{Mullan1992}
Mullan, D.J., Herr, R.B., Bhattacharyya, S., 1992, \apj, 391, 265

\bibitem[Mullan, D.J., et al. 2006]{Mullan2006}
Mullan, D.J., Mathioudakis, M., Bloomfield, D.S., Christian, D.J. 2006, \apjs, in press

\bibitem[Nakariakov \& Ofman 2001]{Nakariakov2001}
Nakariakov, V.M., Ofman, L., 2001, \aap, 327, L53

\bibitem[Nakariakov et al. 2003]{Nakariakov2003}
Nakariakov, V.M., Melnikov, V.F., Reznikova, V.E., 2003, \aap, 412, L7

\bibitem[Nakariakov et al. 2004]{Nakariakov2004}
Nakariakov, V.M., et al., 2004, \aap, 414, L25

\bibitem[Nakariakov \& Verwichte 2005]{Nakariakov2005}
Nakariakov, V. M., Verwichte, E. 2005, Living Rev. Solar Phys., 2, 3 
http://www.livingreviews.org/lrsp-2005-3

\bibitem[Nakariakov et al. 2006]{Nakariakov2006}
Nakariakov, V. M., Foullon, C., Verwichte, E., Young, N.P., 2006, \aap, in press

\bibitem[Roberts, Edwin, \& Benz 1983]{Roberts1983}
Roberts, B., Edwin, P.M., Benz, A.O.,  1983, Nature, 305, 688

\bibitem[Roberts, Edwin, \& Benz 1984]{Roberts1984}
Roberts, B., Edwin, P.M., Benz, A.O.,  1984, \apj, 279, 857

\bibitem[Roberts 2000]{Roberts2000}
Roberts, B. 2000, \solphys, 193, 139

\bibitem[Rodono 1976]{Rodono1976}
Rodon\'{o}, M., 1976, \aap, 32, 337

\bibitem[Saar \& Linsky 1985]{Saar1985}
Saar, S.H., Linsky, J.L., 1985, \apj, 299, L47

\bibitem[Saar, Linsky, \& Beckers 1986]{Saar1986}
Saar, S.H., Linsky, J.L., Beckers, J.M., 1986, \aap, 302, 777

\bibitem[Schrijver, Aschwanden, \& Title 2002]{Schrijver2002}
Schrijver, C.J., Aschwanden, M.J., Title, A.M., 2002, \solphys, 206, 69		

\bibitem[Stepanov et al. 2005]{Stepanov2005}
Stepanov, A.V., Kopylova, Y.G., Tsap, Y.T., Kupriyanova, E.G., 2005, Astron. Lett. 31, 612

\bibitem[Tajima et al. 1987]{Tajima1987}
Tajima, T., et al., 1987, \apj, 321, 1031

\bibitem[Taroyan et al. 2005]{Taroyan2005}
Taroyan, Y., Erd\'{e}lyi, R., Doyle, J.G., Bradshaw, S.J., 2005, \aap, 438, 713

\bibitem[Torrence \& Compo 1998]{Torrence1998}
Torrence, C., Compo, G.P., 1998, Bull. Am. Meteor. Soc. 79, 61

\bibitem[Tsiklauri et al. 2004]{Tsiklauri2004}
Tsiklauri, D., Aschwanden, M.J., Nakariakov, V.M., Arber, T.D., 2004, \aap, 419, 1149

\bibitem[Zhilyaev et al. 2000]{Zhilyaev2000}
Zhilyaev et al. 2000, \aap, 364, 641 
%\end{thebibliography}

%%%%
\end{document}